\documentclass[12pt]{article}
\usepackage{amsmath, amsthm, amscd, amsfonts, amssymb, graphicx, color}
\usepackage[bookmarksnumbered, colorlinks, plainpages]{hyperref}
\usepackage{amsmath,latexsym,amssymb}
\hypersetup{colorlinks=true,linkcolor=red, anchorcolor=green, citecolor=cyan, urlcolor=red, filecolor=magenta, pdftoolbar=true}
\newtheorem{theorem}{Theorem}[section]
\newtheorem{lemma}{Lemma}[section]

\theoremstyle{definition}
\newtheorem{definition}{Definition}[section]

\numberwithin{equation}{section}

\begin{document}

\setcounter{page}{1}

	\thispagestyle{empty}
\renewcommand{\thefootnote}{\fnsymbol{footnote}}

\sloppy

\begin{center}
	{\noindent\bf\Large {Semiconformal symmetry- A new symmetry of the spacetime manifold of the general relativity}}\\
	\vspace{.5cm}
	{\noindent\small{  MUSAVVIR ALI$^{1}\footnote{Corresponding author}$, NAEEM AHMAD PUNDEER$^2$ AND ZAFAR AHSAN$^{3}$ }}\\
	
\end{center}
\vspace{1cm}

	\textbf{Abstract.} \parindent=8mm In this paper we have introduced a  new symmetry property of spacetime which is named as semiconformal curvature collineation, and its relationship with other known symmetry properties has been established. This new symmetry property of the spacetime has also been studied for non-null and null electromagnetic fields.\\

\vspace{.01mm}

\section{Introduction }\label{sec1}
	\noindent In recent years general relativists have been much interested in symmetries of spacetime in general relativity. Such interest is due to the need to simplify the Einstein's field equations in the search
for their exact solutions. These geometrical symmetries of the spacetime are often defined by the vanishing of the Lie derivative of certain tensors with respect to a vector ( this vector may be time-like, space-like or null). The  symmetries in general theory of relativity have been introduced by Katzin, Levine and Davis in the papers (\cite{Katzin curv Collin: A Funda.} and \cite{katzin3}). These symmetries which are also known as collineations, were further studied by Ahsan (\cite{AZ collineation in electromagnetic} - \cite{AZAN}), Ahsan and Hussain (\cite{AZ Hussain}), Ahsan and Siddiqui (\cite{Ahsan Siddiqi Concircular curvature tensor}), Ahsan and Ali (\cite{ahsan and ali type d} - \cite{AZMA} ) and Ali and Ahsan (\cite{Ali M and Ahsan soliton and symm}-\cite{Ali M and Ahsan quasi conformal}) among many others. However, in this paper our study is focused on these symmetries which can be used as simplifying assumptions in the exact solution of Einstein's field equations but solving EFE by our findings will be the next target. Main objective of this paper is to give new symmetry in mathematical approach and analyse it on parameters of the well established literature on symmetries of spacetime manifolds. \\
 As a special subgroup of the conformal transformation group, Ishii(\cite{Ishii Y On Conharmonic Transformation}) defined a rank four tensor $L^h_{bcd}$ that remains invariant under conharmonic transformation for a n-dimensional Riemannian differentiable manifold ($M^n, g$) of dimension $n \geq 4$, as follows:

 \begin{equation}\label{eq1.1}
 L^h_{bcd}=R^h_{bcd}+\dfrac{1}{n-2}(\delta^h_{c}R_{bd} -\delta^h_{d}R_{bc} + g_{bd}R^h_{c} - g_{bc}R^h_{d}),
 \end{equation}

 \noindent where $R^h_{bcd},~R_{bd}$ are Riemann curvature tensor and Ricci tensor respectively. The geometric properties of conharmonic curvature tensor have been discussed by Shaikh and Hui ( \cite{AA Shaikh Hui sk On Weakly Conharmonically} ), while the relativistic significance of this tensor has been investigated by Abdussattar and Dwivedi (\cite{Abdussattar DB On conharmonic trnsf. in GTR}) and Siddiqui and Ahsan (\cite{Siddiqui Zafar conharmonic curvature}). In 2016, J. Kim (\cite{kim Jaemann : A Type of conformal curvature tensor }) introduced curvature like tensor field which remain invariant under conharmonic transformation. He named this new tensor as semiconformal curvature tensor and denoted it by  $P^h_{bcd}$. For a Riemannian manifold $M^n$ with metric $g$, this tensor is defined as (see also \cite{Kim Jaeman On Pseudo Semiconformally })

 \begin{equation}\label{eq1.2}
 P^h_{bcd}=-(n-2)BC^h_{bcd}+[A+(n-2)B]L^h_{bcd},
 \end{equation}

 \noindent provided the constants A and B are not simultaneously zero, $C^h_{bcd}$ is conformal curvature tensor defined as

 \begin{equation}\label{eq1.3}
 C^h_{bcd}=R^h_{bcd}+\dfrac{1}{n-2}(\delta^h_{c}R_{bd} -\delta^h_{d}R_{bc} + g_{bd}R^h_{c} - g_{bc}R^h_{d} )+\dfrac{R}{(n-1)(n-2)}(\delta^h_{d}g_{bc}-\delta^h_{c}g_{bd}),
 \end{equation}
	where $R_{ab}$ is the Ricci tensor and $R$ is the scalar curvature.\\
	
	 For a special substitution $A=1$ and $B=\dfrac{-1}{(n-2)},$ the semiconformal curvature tensor reduces to conformal curvature tensor, while for $A=1$ and $B=0$, it reduces to conharmonic curvature tensor. The semiconformal curvature $P_{hbcd}$ satisfy the following symmetry properties
	
	 \begin{equation}\label{eq1.4}
	 P_{hbcd}=-P_{bhcd}=-P_{hbdc}=P_{cdhb},
	 \end{equation}
	
	 and
	
	 \begin{equation}\label{eq1.5}
	 P_{hbcd}+P_{chbd}+P_{bchd}=0.
	 \end{equation}
	 \noindent In this paper we define a new symmetry in terms of semiconformal curvature tensor and study its relationship with other symmetries of the spacetime. We call this new symmery as semiconformal curvature collineation. Section 2 contains some known results that are required for our investigation. In sections 3 and 4, the relationship between semiconformal cuvature collineation and the other symmetry properties for a general Riemannian space and for a Riemannian space with vanishing Ricci Tensor, respectively, have been established. Finally in section 5, the semiconformal curvature collineation has been studied for non-null and null electromagnetic fields.\\
	
	 \section{Preliminaries}\label{sec2}
	
	 \noindent A geometrical symmetry of the spacetime is often defined in terms of the Lie derivative of a tensor. These symmetries are also known as collineations. The  literature on such collineations is very large and still expanding with results of elegance. Here we shall mention only those symmetry assumptions that are necessary for our study and we have (cf. \cite{Eisenhart LP}, \cite{Schouten JA}, \cite{Yano K The Theory of Lie Derivatives})\\
	
	 \begin{definition}\label{def2.1}
	 	{\it { Motion (M)}} A spacetime is said to admit motion if there exists a vector field $\xi^a$ such that\footnote{Indices takes the values from $1,2,3......,n$ and the summation convention is used. Covariant differentiation is indicated by a semicolon (;) and partial differentiation by a comma (,).}
	 \end{definition}
 \begin{equation}\label{eq2.1}
 {\Huge \eta}_{ab} \equiv \pounds _\xi g_{ab}=\xi _{a;b}+\xi _{b;a}=0.
 \end{equation}
 	Equation (2.1) is known as Killing equation and the vector $\xi^a$ is known as a Killing vector field.\\
 	
 	\begin{definition}\label{def2.2}
 		 {\it Affine Collineation (AC)} A spacetime is said to be an affine collineation if there is a vector field $\xi^a$ such that
 	\end{definition}
	
	 \begin{equation}\label{eq2.2}
	 \pounds_\xi \begin{Bmatrix}
	 c\\ab
	 \end{Bmatrix} \equiv \dfrac{1}{2}g^{cd}({\Huge \eta}_{da;b}+{\Huge \eta}_{db;a}-{\Huge \eta}_{ab;d})=0,
	 \end{equation}
	
	 	where $\begin{Bmatrix}
	 c\\ab
	 \end{Bmatrix}$ is the Christoffel symbol of second kind. Hence the necessary and sufficient condition for an AC [from \eqref{eq2.2}] is
	
	 \begin{equation}\label{eq2.3}
	 {\Huge \eta}_{ab;c}=0,
	 \end{equation}
	
	 	It may be noted, from equations \eqref{eq2.2} and \eqref{eq2.3}, that every M is AC.\\
	 \begin{definition}\label{def2.3}
	 	{\it  Conformal Motion (Conf M)} A spacetime is said to admit a conformal motion if there exist a vector field $\xi^a$ such that
	 \end{definition}

 \begin{equation}\label{eq2.4}
 \pounds_{\xi} \mathfrak{K}_{ab}=0,
 \end{equation}
	
	 where $\mathfrak{K}$ (\cite{Schouten JA}) is defined by
	
	 \begin{equation}\label{eq2.5}
	 \mathfrak{K}=g^{-\frac{1}{4}}g_{ab}.
	 \end{equation}
	\noindent	Accordingly we have
	
	\begin{equation}\label{eq2.6}
	\eta_{ab}=2\phi g_{ab},
	\end{equation}
	\noindent	where $\phi$ is scalar and we may express in the following form
	
	 \begin{equation}\label{eq2.7}
	 \phi=\dfrac{1}{4} \xi^d_{;d}.
	 \end{equation}
	
	\begin{definition}\label{def2.4}
	{\it Projective Collineation (PC)} A spacetime is said to admit projective collineation if there exist a vector $\xi^a$ such that
	\end{definition}
	
	\begin{equation}\label{eq2.8}
	\pounds_\xi \prod_{bc}^{a}=0,
	\end{equation}	
		where the projective connection is defined as for $n=4$
	
	\begin{equation}\label{eq2.9}
	\prod_{bc}^{a}= \begin{Bmatrix} a\\bc \end{Bmatrix}-\dfrac{1}{5} \Bigg(\delta ^a_{b}\begin{Bmatrix} h\\hc \end{Bmatrix}+\delta^a_{c}\begin{Bmatrix} h\\hb \end{Bmatrix}\Bigg).
	\end{equation}		
	From equations \eqref{eq2.8} and \eqref{eq2.9}, we get	
	
\begin{equation}\label{eq2.10}
\pounds_\xi \begin{Bmatrix} a\\bc \end{Bmatrix}= \delta^a_{b}  \sigma_{;c}+ \delta^a_{c} \sigma_{;b},
\end{equation}

where

\begin{equation}\label{eq2.11}
\sigma_{;c} = \frac{1}{5} \xi^m_{;mc} ~~~~ \mbox{and}~~~~  \sigma_{;b} = \frac{1}{5} \xi^m_{;mb}.
\end{equation}
	Further, for every projective collineation, we have (\cite{Katzin curv Collin: A Funda.})
	
\begin{equation}\label{eq2.12}
\pounds_\xi W^h_{bcd}=0,
\end{equation}	

where the Weyl projective curvature tensor for $n=4$ is given by		
		
\begin{equation}\label{eq2.13}
W^h_{bcd}=R^h_{bcd}-\frac{1}{3}(\delta^h_{d} R_{bc} - \delta^h_{c} R_{bd}).
\end{equation}	

\noindent From equations \eqref{eq2.10}, \eqref{eq2.11} and \eqref{eq2.12} it follows that every AC is PC.\\	

\begin{definition}\label{def2.5}
	{\it Conformal Collineation  (Conf C)} A spacetime is said to admit a conformal collineation if there exist a vector $\xi^a$ such that
\end{definition}	

\begin{equation}\label{eq2.14}
\pounds_\xi \begin{Bmatrix}
a\\bc
\end{Bmatrix} = \delta^a_{b} \phi_{;c} + \delta^a_{c} \phi_{;b} - g_{bc}g^{am} \phi_{;m},
\end{equation}	
		
where 	$\phi = \frac{1}{4} \xi^d_{;d}$.

\noindent\\
	\noindent Equations \eqref{eq2.6} and \eqref{eq2.14} may be expressed as (\cite{Katzin curv Collin: A Funda.})
	
	\begin{equation}\label{2.15}
	\eta_{ab;c} = 2\phi_{;c} g_{ab},
	\end{equation}
	\noindent	and that every Conf C  must satisfy (for explanation c.f., \cite{Yano K The Theory of Lie Derivatives})
	
	\begin{equation}\label{eq2.16}
	\pounds_\xi C^h_{bcd} = 0,
	\end{equation}
	where $C^h_{bcd}$ is conformal curvature tensor, which  from equation \eqref{eq1.3} for $n=4$, is given by	
	\begin{equation}\label{eq2.17}
	C^h_{bcd}=R^h_{bcd}+\frac{1}{2}( \delta^h_{c}R_{bd} - \delta^h_{d}R_{bc}+g_{bd}R^h_{c} - g_{bc}R^h_{d})+\frac{R}{6}(\delta^h_{d}g_{bc}-\delta^h_{c}g_{bd} ).
	\end{equation}	

\begin{definition}\label{def2.6}
	{\it Curvature Collineation (CC)} A spacetime is said to admit curvature collineation if there is a vector field $\xi^a$ such that
\end{definition}
\begin{equation}\label{eq2.18}
\pounds_{\xi} R^h_{bcd}=0,
\end{equation}	
	
	\noindent where Riemann curvature tensor is defined as (\cite{Eisenhart LP})
	
\begin{equation}\label{eq2.19}
R^h_{bcd}=\begin{Bmatrix}
h\\bd
\end{Bmatrix}_{,c} - \begin{Bmatrix}
h\\bc
\end{Bmatrix}_{,d} + {\begin{Bmatrix}
	m\\bd
	\end{Bmatrix}}{ \begin{Bmatrix}
	h\\mc
	\end{Bmatrix}} - {\begin{Bmatrix}
	m\\bc
	\end{Bmatrix}} {\begin{Bmatrix}
	h\\md
	\end{Bmatrix}}.
\end{equation}		
	
\begin{definition}\label{def2.7}
	{\it Ricci Collineation (RC)} A spacetime is said to admit Ricci collineation if there is a vector field $\xi^a$ such that
\end{definition}

\begin{equation}\label{eq2.20}
\pounds_\xi R_{ab}=0,
\end{equation}

\noindent	where $R_{ab}$ is the Ricci tensor.

\begin{definition}\label{def2.8}
	{\it Maxwell collineation (MC)} The electromagnetic field inherits the symmetry property of spacetime such that
\end{definition}

\begin{equation}\label{eq2.21}
\pounds_\xi F_{ab}=F_{ab;c}\xi ^c+F_{ac}\xi _{;b}^c+F_{bc}\xi _{;a}^c=0,
\end{equation}

\noindent	where $F_{ab}$ is the electromagnetic field tensor. A point transformation that leave $F_{ab}$ invariant, i.e., equation (2.21) is satisfied, is called a Maxwell collineation (\cite{Collinson CD General Relativity and Gravitation}).\\

\section{Semiconformal Symmetry}\label{sec3}

\noindent  For $n=4$, the semiconformal curvature tensor, from equation \eqref{eq1.2}, is given by
\begin{equation}\label{eq3.1}
P^h_{bcd}=-2BC^h_{bcd}+[A+2B]L^h_{bcd},
\end{equation}

\noindent	where $C^h_{bcd} $ and $L^h_{bcd}$ are the conformal and conharmonic curvature tensor respectively.\\

 We now define a new symmetry for the spacetime manifold of general relativity as

\begin{definition}\label{def3.1}
{\it Semiconformal Curvature Collineation (Semiconf CC)} A spacetime $V_{4}$ is said to admit a semiconformal curvature collineation if there exists a vector field $\xi^a$ such that
\end{definition}
\begin{equation}\label{eq3.2}
\pounds_\xi P^h_{bcd}=0,
\end{equation}
\noindent	where $P^h_{bcd}$ is semiconformal curvature tensor is defined in equation \eqref{eq1.2}.

\begin{definition}\label{def3.2}
	{\it Special  Semiconformal Curvature Collineation (S Semiconf CC)} A semiconformal curvature collineation with the following condition
\end{definition}
\begin{equation}\label{eq3.3}
\sigma_{;bc} = 0,
\end{equation}
\noindent is called a special semiconformal curvature collineation.

 \noindent where $\sigma_{;bc} = \dfrac{1}{4}\xi^d_{;dbc}$ and $\xi^a$ is Killing vector field.\\

\noindent We also define

\begin{definition}\label{def3.3}
	 {\it Conharmonic Curvature Collineation (Conh CC)} A spacetime $V_4$ is said to admit a conharmonic curvature collineation if there exists a vector $\xi^a$ such that
\end{definition}
\begin{equation}\label{eq3.4}
\pounds_\xi L^h_{bcd}=0,
\end{equation}

\noindent	where conharmonic curvature tensor $L^h_{bcd}$ is defined by (\cite{Ahsan Z Tensors Mathematics of differential }, \cite{Siddiqui Zafar conharmonic curvature} )

\begin{equation}\label{eq3.5}
L^h_{bcd} = R^h_{bcd}+\frac{1}{2}( \delta^h_{c}R_{bd} - \delta^h_{d}R_{bc}+g_{bd}R^h_{c} - g_{bc}R^h_{d}).
\end{equation}

\begin{definition}\label{def3.4}
	 {\it {Concircular Curvature Collineation (Conc CC)}} A spacetime $V_4$ is said to admit a concircular curvature collineation if there exists a vector $\xi^a$ such that
\end{definition}

\begin{equation}\label{eq3.6}
\pounds_\xi M^h_{bcd} = 0,
\end{equation}

\noindent	where concircular curvature tensor $M^h_{bcd}$ is defined by (\cite{Ahsan Z Tensors Mathematics of differential }, \cite{Ahsan Siddiqi Concircular curvature tensor})

\begin{equation}\label{eq3.7}
M^h_{bcd} = R^h_{bcd} - \dfrac{R}{12}(\delta^h_{d}g_{bc}-\delta^h_{c}g_{bd}).
\end{equation}
\begin{center}
 \bf Main results of the section 3.
\end{center}		
 \begin{theorem}\label{th3.1}
 The necessary and sufficient condition for a semiconformal curvature collineation (Semiconf CC)  to be a curvature collineation (CC) is that

 \begin{equation*}
  \phi_{;bc} = 0,
 \end{equation*}

 \noindent where $\phi = \dfrac{1}{4} \xi^d_{;d}$ is a scalar function.
\end{theorem}

\begin{theorem}\label{th3.2}
	The necessary and sufficient condition for a semiconformal curvature collineation to be a Weyl projective curvature collineation is that
	
	\begin{equation*}
	\sigma_{;bc}=0.
	\end{equation*}
	\noindent	where $\sigma = \dfrac{1}{5}\xi^m_{;m}.$\\
\end{theorem}

\begin{theorem}\label{th3.3}
	A spacetime $V_4$ admits semiconformal curvature collineation along a vector field ${\xi}^a$ provided that ${\xi}^a$ is Killing.
\end{theorem}

\section{Semiconformal symmetry in empty spacetime}
\noindent The Einstein field equations are given by

\begin{equation}\label{eq4.1}
R_{bc} - \dfrac{1}{2}g_{bc}R=-kT_{bc}
\end{equation}
\noindent where $R_{bc}$ is the Ricci tensor, $g_{bc}$ is the metric tensor, $T_{bc}$ is the energy momentum tensor, $R$ is the scalar curvature tensor and $k$ is the constant.\\
\noindent Multiplying by$g^{bc}$ and using $g^{bc}g_{bc}=4$, equation \eqref{eq4.1} takes the form
\begin{equation}\label{eq4.2}
R=kT.
\end{equation}
\noindent from the equations \eqref{eq4.1} and \eqref{eq4.2}, we get

\begin{equation}\label{eq4.3}
R_{bc} = k(T_{bc}- \dfrac{1}{2}g_{bc}T)
\end{equation}
\noindent If $T_{bc}=0$, then $T=g^{bc}T_{bc}=0$, equation \eqref{eq4.3} yields

\begin{equation}\label{eq4.4}
R_{bc}=0,
\end{equation}
\noindent these equations are the field equations for empty spacetime.
\begin{center}
 \bf Main results of the section 4.
\end{center}
\begin{theorem}\label{th4.1}
	In an Empty spacetime $V^0_4 $ Lie derivative of semiconformal curvature tensor is proportional to Lie derivative of Riemann curvature tensor.
\end{theorem}

\begin{theorem}\label{th4.2}
	In empty spacetime $V^0_4 $ the Lie derivatives of semiconformal curvature and Weyl projective curvature tensors are proportional.
\end{theorem}

\section{Semiconformal collineation and electromagnetic fields}
	\noindent It is known that in general relativity, the electromagnetism can be described through Maxwell's equation
	
	\begin{equation}\label{eq5.1}
	F_{[ab;c]} = 0,~~~~ F^{ab}_{;b} = J^a,
	\end{equation}
	\noindent where the skew-symmetric tensor $F_{ab}$ represents the electromagnetic field tensor and $J^a$ the current density. Moreover, we have defined the Einstein field equations in equation \eqref{eq4.1} and in presence of matter in  equation \eqref{eq4.3}\\
\noindent The energy - momentum tensor for an electromagnetic field is given by
\begin{equation}\label{eq5.2}
T_{ab} = -F_{ac}F^c_b + \dfrac{1}{4}g_{ab}F_{pq}F^{pq},
\end{equation}	
\noindent which is symmetric tensor. Equation \eqref{eq5.2} leads to $T^a_a = T = 0$ and thus the Einstein equation for a purely electromagnetic distribution is given by

\begin{equation}\label{eq5.3}
R_{ab} = k T_{ab},
\end{equation}

\noindent The geometrical symmetry defined by equation \eqref{eq2.21} along with the symmetry given by equation \eqref{eq2.1} has been the subject of interest for quite some time. Thus for example, for non-null electromagnetic fields Woolley (\cite{Woolley MC The structure of groups}) has shown that if equation \eqref{eq2.1} holds then $F_{ab}$ satisfies $\pounds_{\xi} F_{ab} = k(\alpha)F_{ab}$ form some constant $k(\alpha),$ $\alpha = 1,2,..........r;$ while Michalski and Wainwright (\cite{Michalaski H Wainwright j. Killing vector Fields}) have shown that $\pounds_{\xi} g_{ab} = 0 \implies \pounds_{\xi} F_{ab} = 0$ for non-null fields. On the other hand, for non-null fields, Duggal (\cite{Duggal- Existence of two Killing}) has proved the converse part under certain conditions. Maxwell collineations have also been studied by Ahsan and Ahsan and Husain (\cite{AZ Hussain}, \cite{AZ collineation in electromagnetic}, \cite{AZ Symmetries of the Electrom}). It is seen that for null electromagnetic fields neither MC is a consequence of Motion nor Motion is a consequence of Maxwell collineation. Moreover, using Newman-Penrose formalism, Ahsan (\cite{AZ Symmetries of the Electrom}) has obtained the conditions under which a null electromagnetic field may admit Maxwell collineation and Motion. The concept of Maxwell collineation was further extended as Maxwell Inheritance (MI) by Ahsan and Ahsan (\cite{AZAN}), who applied this concept to (i) the spacetime solution corresponding to strong gravitational waves propagating in generalised electromagnetic universes and (ii) the algebraically general twist-free solution of Einstein-Maxwell equation for non-radiative electromagnetic fields.

In 1986 Khlebnikov (\cite{Khleb}) has obtained the solutions for Einstein-Maxwell equations corresponding to the strong gravitational waves in the generalized electromagnetic universe. He used the technique of N.P. formalism (\cite{NP}) to obtain the solution in non-radiative electromagnetic fields. In his solution he took the first real null tetrad vector $l^a$ as geodetic and shear-free  and the tetrad as parallelly propagated along $l^a$ also proved that the solution does not admit the Maxwell collineation. As another {\it example} we can refer the twist-free algebraically general solution given by Tariq and Tupper (\cite{TT}). For this solution of Einstein-Maxwell equations in non-null electromagnetic fields together with the condition of coupling theorem Tariq and Tupper proved that their solutions also do not admit the Maxwell collineation.

Motivated by above discussions, in this section, we shall investigate the role of semiconformal curvature collineation to the non-null and null electromagnetic fields.\\
\begin{lemma}\label{lema5.1}
	(\cite{Michalaski H Wainwright j. Killing vector Fields})  In a non-null electromagnetic field, the Lie derivative of electromagnetic field tensor $F_{ab}$ with respect to a vector field $\xi$ vanishes, if $\xi$ is Killing vector.\\
\end{lemma}

\begin{center}
  \bf Main results of section 5.
\end{center}

\begin{theorem}\label{th5.1}
	 A non-null electromagnetic field admits semiconformal curvature collineation along a Killing vector field.
\end{theorem}

\begin{theorem}\label{th5.2}
	A non-null electromagnetic field admits semiconformal curvature collineation if it admits Maxwell collineation.
\end{theorem}

\begin{theorem}\label{th5.3}
	A null electromagnetic field admits semiconformal curvature collineation along a propagation (polarization) vector if propagation (polarization) vector is killing and expansion-free.
\end{theorem}

\noindent Note: Proof of the main results will be uploaded after publication of the article.

\vspace{1cm}
{\noindent\small {Authors Address$^{1, 2, \&3}$:\\
		DEPARTMENT OF MATHEMATICS\\
		ALIGARH MUSLIM UNIVERSITY\\
		ALIGARH-202002, INDIA\\
		E-MAILS :\\
		1: musavvirali.maths@gmail.com,\\
		2: pundir.naeem@gmail.com. }}\\
3: zafar.ahsan@rediffmail.com,\\		


\begin{thebibliography}{99}
	\bibitem{Abdussattar DB On conharmonic trnsf. in GTR}   Abdussattar and Dwivedi, B.: {\it On Conharmonic transformations in general relativity.} Bull. Calcutta Math. Soc., 88(6) 465-470 (1996).
	
	\bibitem{Ahsan Z Tensors Mathematics of differential }  Ahsan, Z.: {\it Tensors Mathematics of Differential Geometry and Relativity.} PHI Learning Pvt. Ltd., Delhi, (2015).
	
	
	\bibitem{AZ collineation in electromagnetic}  Ahsan, Z.: {\it Collineation in electromagnetic fileds in general relativity-The null field case.} Tamkang Journal of Mathematics, 9(2) 237-240 (1978).
	
	
	\bibitem{AZ On the Nijenhuis tensor} Ahsan, Z.: {\it On the Nijenhuis tensor in null electromagnetic field.} Journal of Math. Phys. Sci., 21(5) 515-526 (1987).
	
	\bibitem{AZ Symmetries of the Electrom} Ahsan, Z.: {\it Symmetries of the electromagnetic fields in general relativity.} Acta Phys. Sinica, 4(5) 337 (1995).
	
	\bibitem{AZ A symmetry property of the spacetime} Ahsan, Z.: {\it A symmetry property of the spacetime of general relativity in terms of space-matter tensor.} Braz. J. Phys., 26(3) 572-576 (1996).
	
	\bibitem{AZ on a geom. bull} Ahsan, Z. : {\it On a geometrical symmetry of the spacetime of general relativity.} Bull. Cal. Math. Soc., 97(3)191-200 (2005).
	\bibitem{AZAN} Ahsan, Z. and Ahsan, N.: {\it On a symmetry of the electromagnetic fields.} Bull. Cal. Math. Soc., 94(5)385-388 (2002).
	
	
	\bibitem{ahsan and ali type d} Ahsan, Z. and Ali, M.: {\it Symmetries of type $D$ pure radiation fields.} Int. Jour. of Theo. Phys., 51 2044-2055 (2012).
	
	\bibitem{ahsan and ali type n} Ahsan, Z. and Ali, M.: {\it ``Symmetries of Type N Pure Radiation Fields".}  Int. Jour. Theo.
	Physics, 54(5) 1397 - 1407 (2015).
	
	\bibitem{ahsan and ali type W} Ahsan, Z. and Ali, M.:{\it ``On some properties of the W-curvature tensor".}  Palestine Journal of
	Mathematics, (ISSN- 2219-5688), 3(1) 61-69 (2014).
	\bibitem{AZMA}  Ahsan, Z. and Ali, M. : {\it Curvature tensor for spacetime of general relativity.} Int. Jour. Geom. Methods Mod. Phys.,  14(5) 1750078(13 pages) (2017)  DOI: 10.1142/S0219887817500785.
	
	\bibitem{AZ Hussain} Ahsan, Z. and Hussain, S.I.: {\it Null electromagnetic fields, total gravitational radiation and collineations in general relativity.} Ann. di Mat. Pur. ed Applicata CXXXVI, 379-396 (1980).
	\bibitem{Ahsan Siddiqi Concircular curvature tensor} Ahsan, Z. and Siddiqui, S.A.: {\it Concircular Curvature Tensor and Fluid spacetime.} Int. Jour. of Theo. Phys., 48 3202-3212 (2009).
	\bibitem{Ali M and Ahsan soliton and symm} Ali, M. and Ahsan, Z.: {\it Ricci solitons and symmetries of spacetime manifold of general relativity.} Global Journal of Advanced Research on Classical and Modern Geometries, 1(2) 76-85 (2012).
	
	\bibitem{Ali and Ahsan soliton and N} Ali, M. and Ahsan, Z.: {\it On Ricci Soliton and symmetries of type N PR fields.} Proceedings, Int. Conf. D.G. Functional Analysis and Applications, JMI, New Delhi, editor - M. Hasan Shahid et al., ISBN: 978-81-8487-421-1, 1-10 (2014).
	
	\bibitem{Ali M and Ahsan quasi conformal} Ali, M. and Ahsan, Z.: {\it Quasi-Conformal Curvature Tensor for the spacetime of general relativity.}  Palestine Journal of Mathematics, ISSN- 2219-5688, 4(1) 233-240 (2015).
	
	\bibitem{Collinson CD General Relativity and Gravitation} Collinson, C.D.:{\it Conservation laws in general relativity based upon existence of preferred collineation.} General Relativity and Gravitation, 1, 137  (1970).
	
	\bibitem{Duggal- Existence of two Killing} Duggal, K. L. : {\it Existence of two Killing vector fieldson the spacetime of general relativity.} (Preprint Tensor New Series), (1978).
	
	\bibitem{Eisenhart LP} Eisenhart L.P.: {\it Riemannian Geometry.} Princeton University Press, Princeton, N.J., (1926).
	
	\bibitem{Hazen MZ Black holes in static} Hagen, M.Z., Robinson, D.C., Seifert, H.J.:{\it Black holes in static electrovac space-times}. General Relativity and Gravitation, 5(1) 61-72 (1974).
	
	\bibitem{Israel W Comm Math Physics} Israel, W.: {\it Event horizons in static electrovac space-times.} Comm. Math. Phys, 8(3) 245-260 (1968).
	
	\bibitem{Ishii Y On Conharmonic Transformation}   Ishii, Y.: {\it On Conharmonic transformations.} Tensor (N.S.), 7, 73-80 (1957).
	
	\bibitem{Katzin curv Collin: A Funda.}  Katzin, G.H., Livine, J. and Davis, W.R.: {\it Curvature collineation: A fundamental symmetry property of the space-times of general relativity defined by the vanishing Lie derivative of the Riemann curvature tensor.} Journal of Mathematical Physics, 10(4) 617-620 (1969).
	
	\bibitem{katzin3}  Katzin, G.H., Livine, J. and Davis, W.R.: Journal of Mathematical Physics, 11 1875 (1970).

\bibitem{Khleb}	Khlebnikov, V. I.: {\it Gravitaional radiation in electromagnetic universes.} Class. Quant. Grav.3(2)(1986).

	\bibitem{kim Jaemann : A Type of conformal curvature tensor } Kim, J.: {\it A type of conformal curvature tensor.} Far East J.  Math. Sci., 99(1) 61-74 (2016).
	
	\bibitem{Kim Jaeman On Pseudo Semiconformally } Kim, J.: {\it On pseudo semiconformally symmetric manifolds.} Bull. Korean Math. Soc., 54(1) 177-186 (2017).
	
	\bibitem{Michalaski H Wainwright j. Killing vector Fields}  Michalaski, H. and Wainwright, J.: {\it Killing vector fields and the Einstein-Maxwell field equations in general relativity.} Genral Relativity and Gravitation, 6(3) 289 (1975).
	
\bibitem{NP} Newman, E.T. and Penrose, R.: {\it An approach to gravitational radiation by a method of spin coefficients.} J. Math. Phys. 3, 566 (1962).

	\bibitem{Siddiqui Zafar conharmonic curvature} Siddiqui, S.A. and Ahsan, Z.: {\it Conharmonic curvature tensor and the spacetime of General Relativity.} Diff. Geom. and Dynamical System, 12, 213-220 (2010).
	
\bibitem{TT} Tariq, N. and Tupper, B.O.J.: {\it A class of algebraically general solutions of the Einstein-Maxwell equations for non-null electromagnetic field.} Class. Quant. Grav. 6, 345 (1975).
	
	\bibitem{AA Shaikh Hui sk On Weakly Conharmonically} Shaikh, A.A. and Hui, S.K.: {\it On Weakly conharmonically symmetric manifolds.} Tensor (N.S.), 70(2) 119-134 (2008).
	
	\bibitem{Schouten JA} Schouten, J.A.: {\it Ricci Calculus (An Introduction to Tensor Analysis and its geometrical Applications).} Springer-Verlag, Berlin, 2nd edition, editor - Courant, R. (Hrsg.) (1954).
	
	\bibitem{Woolley MC The structure of groups} Woolley, M.L.: {\it The structure of groups of motions admitted by Einstein-Maxwell space-times.} Comm. Math. Phys., 31(1) 75-81 (1973).
	
	\bibitem{Yano K The Theory of Lie Derivatives}   Yano, K.: {\it The Theory of Lie Derivatives and its Applications.} North Holand Publishing Co. Amsterdom P. Noordhoff L.T.D. Groningen, Vol. III (1957).
	
\end{thebibliography}
\end{document}